\documentclass[11pt]{article}
\usepackage{graphicx}

\newcommand{\BABARPubYear}    {06}

\newcommand{\BABARConfNumber} {019}
\newcommand{\SLACPubNumber} {11998}

\newcommand{\modea} {$K4\pi$}
\newcommand{\modeb} {$K6\pi$}
\def\deltadso{\ensuremath{\Delta m(\Dsop)}\xspace}
\def\dmmean{\ensuremath{\Delta \mu(\Dsop)}\xspace}
\def\gammadso{\ensuremath{\Gamma(\Dso)}\xspace}
\def\Dsop    {\ensuremath{D^{+}_{s1}}\xspace}
\def\Dsom    {\ensuremath{D^{-}_{s1}}\xspace}
\def\Dso     {\ensuremath{D_{s1}}\xspace}

\newcommand{\kevcc}{\ensuremath{{\mathrm{\,ke\kern -0.1em V\!/}c^2}}\xspace}

\input pubboard/babarsym

\long\def\inst#1{\par\nobreak\kern 4pt\nobreak
    {\it #1}\par\vskip 10pt plus 3pt minus 3pt}

\begin{document}
{\pagestyle{empty}

\begin{flushright}
\babar-CONF-\BABARPubYear/\BABARConfNumber \\
SLAC-PUB-\SLACPubNumber \\
July 2006 \\
\end{flushright}

\par\vskip 5cm

\begin{center}
\Large \bf A precision measurement of the {\boldmath$D_{s1}(2536)^{+}$} meson\\ mass and decay width 
\end{center}
\bigskip

\begin{center}
\large The \babar\ Collaboration\\
\mbox{ }\\
\today
\end{center}
\bigskip \bigskip

\begin{center}
\large \bf Abstract
\end{center}
The decay width and the mass of the $\ensuremath{D_{s1}}\xspace(2536)^{\pm}$ have been measured via the
decay channel \mbox{$\ensuremath{D^{\pm}_{s1}}\xspace \to \ensuremath{D^{*\pm}}\xspace \ensuremath{K^0_{\scriptscriptstyle S}}\xspace$} using $232\ensuremath{\mbox{\,fb}^{-1}}\xspace$ of data collected with the \mbox{\slshape B\kern-0.1em{\smaller A}\kern-0.1em B\kern-0.1em{\smaller A\kern-0.2em R}} detector at the PEP-II asymmetric-energy $e^{+}e^{-}$ storage ring. The result for the decay
width is $\Gamma(\ensuremath{D^{\pm}_{s1}}\xspace) = (1.03 \pm 0.05 \pm 0.12)\ensuremath{{\mathrm{\,Me\kern -0.1em V\!/}c^2}}\xspace$, with the first error denoting the statistical uncertainty and the second one the systematic uncertainty. For the mass, a value of $m(\ensuremath{D^{\pm}_{s1}}\xspace) = (2534.85 \pm 0.02 \pm 0.40)\ensuremath{{\mathrm{\,Me\kern -0.1em V\!/}c^2}}\xspace$ has been obtained. The systematic error is dominated by the uncertainty on the $D^{\ast\pm}$ mass. The mass difference between the $ \ensuremath{D^{\pm}_{s1}}\xspace $ and $\ensuremath{D^{*\pm}}\xspace$ has been measured to be $\Delta m = (524.85 \pm 0.02 \pm 0.04)\ensuremath{{\mathrm{\,Me\kern -0.1em V\!/}c^2}}\xspace$.

\vfill
\begin{center}

Submitted to the 33$^{\rm rd}$ International Conference on High-Energy Physics, ICHEP 06,\\
26 July---2 August 2006, Moscow, Russia.

\end{center}

\vspace{1.0cm}
\begin{center}
{\em Stanford Linear Accelerator Center, Stanford University, 
Stanford, CA 94309} \\ \vspace{0.1cm}\hrule\vspace{0.1cm}
Work supported in part by Department of Energy contract DE-AC03-76SF00515.
\end{center}

\newpage
} 

\begin{center}
\small

The \babar\ Collaboration,
\bigskip

%
{B.~Aubert,}
{R.~Barate,}
{M.~Bona,}
{D.~Boutigny,}
{F.~Couderc,}
{Y.~Karyotakis,}
{J.~P.~Lees,}
{V.~Poireau,}
{V.~Tisserand,}
{A.~Zghiche}
\inst{Laboratoire de Physique des Particules, IN2P3/CNRS et Universit\'e de Savoie,
 F-74941 Annecy-Le-Vieux, France }
% ...

\end{center}\newpage

\section{INTRODUCTION}
\label{sec:Introduction}
The discovery of the mesons $D^{*}_{sJ}(2317)^{+}$ and $D_{sJ}(2460)^{+}$ \cite{ref:dsjbb,ref:dsjcl}, with masses considerably lower than predicted by potential models \cite{ref:gois,ref:dipe}, has renewed experimental and theoretical interest in the spectroscopy of charmed mesons. For a complete understanding of the charmed strange meson spectrum, a comprehensive knowledge of the parameters of all known \Ds mesons is mandatory.
In this analysis, a precision measurement of the mass and the decay width of the meson $\Dso(2536)^{+}$ has been performed. The mass is currently reported by the PDG with a precision of $0.6 \mevcc$, while only an upper limit of $2.3 \mevcc$ is given for the decay width \cite{ref:pdg2004}. These values are based on measurements with 20 times fewer reconstructed \Dsop candidates compared to this analysis. The \babar\ experiment, in addition to its excellent tracking and vertexing capabilities, provides a rich source of charmed hadrons, enabling an analysis of the \Dsop with high statistics and small errors.

Since the uncertainty of the \Dstarp mass is large ($0.4 \mevcc$ \cite{ref:pdg2004}), we perform a measurement of the mass difference defined by
\begin{equation}
\deltadso = m(\Dsop) - m(\Dstarp) - m(\KS).
\end{equation}
Additionally, due to the correlation between the masses, the \Dsop signal in the mass difference spectrum is much more narrow than the one from the \Dsop mass spectrum alone.

\section{THE \babar\ DETECTOR AND DATASET}
\label{sec:babar}
The data sample used in this analysis corresponds to $232 \invfb$ collected with the \babar\ detector
at the \pep2\ storage ring from $\epem$ collisions at or just below the \FourS resonance. 
Furthermore, $1.16$ million \Dsop Monte Carlo events were generated for each of the two
decay modes which are used for the determination of the detector resolution model. Finally, for resolution studies, \Dz and \KS samples were analyzed using $5\%$ of the main data sample and $20$ million simulated $\epem \to \ccbar$ generic Monte Carlo continuum events.   

The \babar\ detector is described elsewhere~\cite{ref:babar} in detail.
Charged particles are detected with a combination of a five-layer double-sided silicon vertex tracker (SVT) and a 40-layer drift chamber (DCH) filled with a mixture of helium and isobutane, both embedded in the $1.5\,T$ solenoidal magnetic field. The transverse momentum resolution is approximately $\sigma_{p_{t}}/p_{t} = 0.0013(p_{t}/GeV/c) \oplus 0.0045$. Charged particle identification is done via energy loss measurement within the SVT and DCH, and via the Cherenkov light detected in a ring imaging Cherenkov detector (DIRC). Photons are detected with a CsI(Tl) electromagnetic calorimeter consisting of $6580$ CsI(Tl) crystals. The instrumented flux return (IFR) contains resistive plate chambers for the identification of muons and long-lived neutral hadrons. 
For the event simulation we use the Monte Carlo generator EVTGEN~\cite{ref:evgen}
with a full detector simulation that uses GEANT4~\cite{ref:geant}.

The most critical aspect of this analysis is the quality of the track reconstruction. The track-finding algorithm is based on tracks found by the trigger system and by standalone track reconstruction in the SVT and DCH. The parameters of a given track are determined using a Kalman filter algorithm~\cite{ref:kal}, which makes optimal use of the hit information and corrects for energy loss and multiple scattering in the material traversed and for inhomogeneities in the magnetic field. The material-traversal corrections change the track momentum according to the expected average energy loss and increase the covariance for track parameters to account for both multiple scattering and the variance in the energy loss. The latter depends on the particle velocity, so each track fit is performed separately for five particle hypotheses: electron, muon, pion, kaon and proton. A simplified model of the \babar\ detector material distribution is used in the Kalman filter algorithm; a wrong simulation of the material will result in a wrong energy loss correction in the reconstruction. Besides a good description of the tracking region, a detailed knowledge of the magnetic field is also essential for a precise track reconstruction. The solenoid field itself has been well measured. The field strength in the tracking volume is known to an accuracy of $0.2\,\mathrm{mT}$.  More uncertain is the contribution of the solenoid field to the magnetization of the permanent magnets used for final focusing and bending of the beams.     
This is discussed in more detail in section \ref{ssec:detcon}.

\section{ANALYSIS METHOD}
\label{sec:Analysis}
\subsection{{\boldmath $\Dsop$} candidate reconstruction}
\label{ssec:evsel}
Two decay modes are reconstructed: in both modes the \Dsop decays to $\Dstarp\!\KS$, with the \KS decaying into \pipi and the \Dstarp into $\Dz\!\pip$. The \Dz decays either into $\Km\!\pip$ or $\Km\!\pip\!\pip\!\pim$. In the following, we refer to these two decay modes as $K4\pi$ and $K6\pi$, respectively, where also the charge conjugated states are included.  

For the \Dz candidate reconstruction, charged kaon and pion candidates are combined to form $\Km\!\pip$ ($\Km\!\pip\!\pip\!\pim$) final states. 
Fits to the \Dz mass spectra yield a mean value of $1863.6\mevcc$ and a signal
resolution of $7.4\mevcc$ ($8.1\mevcc$) for
the \modea\ (\modeb) decay mode. The signal region is
chosen as a mass window of $\pm18\mevcc$ ($\pm14\mevcc$) centered around the obtained mean value.
The \Dz candidates are each combined with an additional charged pion to form \Dstarp candidates.
The fits to the \Dstarp-\Dz mass difference yield a mean value of $145.4\mevcc$ and a signal resolution
of $0.19\mevcc$ ($0.24\mevcc$) for \modea\ (\modeb). The signal
region is chosen as a mass window of $\pm1.5\mevcc$ centered around
the derived mean value for both decay modes.
Finally, \KS candidates are created from oppositely charged tracks.
From the fits to the \KS mass distributions one obtains a mean value of $497.2\mevcc$  and a resolution of $2.5\mevcc$ for both decay modes. For the further analysis we define the \KS signal region as a $\pm10\mevcc$ mass window centered around the signal mean value for both modes.

The background of the \KS spectrum is further reduced by restricting the angle between the \KS direction of flight and the line connecting the primary vertex and the \KS vertex to values smaller than $0.15\,\mathrm{rad}$. The \Dstarp candidates are finally combined with the \KS candidates to form \Dsop candidates. In order to suppress combinatorial background, a momentum $p^{*} > 2.7 \gevc$ in the center of mass system (CMS) is required for \Dsop candidates. In addition, this restricts the source of the $\Dsop$ candidates to $\epem \to \ccbar$ continuum production. A kinematic fit is applied to the \Dsop candidates which satisfy the above selection criteria in such a way, that the tracks for each composed particle have the same origin. The position of the \Dsop vertex is required to be consistent with the $\epem$ interaction region. Since the mass difference \deltadso is measured, no mass constraint is applied. The probability of the vertex fit is required to be greater than $0.1\%$.  
Initially, more than one \Dsop candidate per event is reconstructed. Although the multiple use of tracks within one reconstructed decay tree is excluded, there might be multiple \Dsop candidates sharing the same daughter candidate. After applying all
selection criteria, the average multiplicity of \Dsop candidates per event is $1.008$ and $1.02$ for the two decay modes.  
The selection efficiency is $16\%$ for the \modea\ decay mode and $11\%$ for the \modeb\ mode.

The resulting mass difference spectra \deltadso for MC and data are shown in Section~\ref{sub:val} and~\ref{sec:ffit}, respectively. 
A double Gaussian is fitted to the data spectrum as a rough estimate of the width of the signal. The results for the total width, calculated by adding the weighted Gaussian widths in quadrature, are $1.55 \pm 0.15\mevcc$ and $1.66 \pm 0.26\mevcc$ for mode \modea\ and \modeb, respectively. Note that for this preliminary fit the intrinsic width and the resolution have not been taken separately into account.

\subsection{Resolution model}
\label{ssec:resmod}
Although very clean signals with more than 2400 (2900) entries have been obtained for decay mode \modea\ (\modeb), it is not feasible to obtain the resolution model and the intrinsic width from a single fit to data with all parameters allowed to vary. Instead a resolution model is derived from the corresponding \Dsop Monte Carlo samples. 
The generation of $\epem \to \ccbar$ fragmentation events with high accuracy is a difficult task, so deviations between simulated and real data are possible. In particular the distribution of the CMS momentum $p^{*}$ of the reconstructed \Dsop mesons differs between data and Monte Carlo events.
In order to extract a reliable resolution model from Monte Carlo, the model is determined as a function of $p^{*}$. Another method to compensate for the inaccuracies of the simulation is to weight the Monte Carlo spectrum according to the $p^{*}$ distribution obtained from data and extract the resolution model from the weighted Monte Carlo sample. Since the second method relies on both real data and Monte Carlo data and is sensitive to the chosen $p^{*}$ binning, it is used as a systematic check and the resolution model is derived from the first method.  

The resolution can be extracted by calculating the difference of the invariant mass of a reconstructed candidate and the corresponding generated mass. The derived distribution consists only of the deconvolved resolution part of the \Dsop signal.
Since for the measurement of the \Dsop
mass the mass difference \deltadso is taken, the mass difference $\Delta m_g(\Dsop)$ of
the corresponding generated candidates has to be subtracted to obtain
the deconvolved distribution:
\begin{equation}
\Delta m_{res}= \deltadso-\Delta m_g(\Dsop).
\end{equation}

The procedure is the same for both decay modes. The Monte Carlo sample is divided into 25 $p^{*}$ bins with a width of $0.07 \gevc$ in the range from $2.70\gevc$ to $4.45\gevc$. An unbinned maximum likelihood fit of $\Delta m_{res}$ is performed for each bin of $p^{*}$ using a probability density function (PDF) assembled from Gaussian functions with fit parameters $\Delta m_{res,0}$, $r$ and $\sigma_0$:
\begin{equation}
  R(\Delta m_{res})=\int_{\sigma_0}^{r\sigma_0} \frac{1}{r\sigma^2}
  e^{-\frac{(\Delta m_{res}-\Delta m_{res,0})^2}{2\sigma^2}}\mathrm{d}\sigma,
\label{eq:iares}
\end{equation}
where $\Delta m_{res,0}$ is the mean value and the width is integrated from a minimum value of $\sigma_0$ up to a maximum of $r\sigma_0$. The scale parameter for the upper width limit $r$ is determined for each $p^{*}$ bin, but does not vary drastically with $p^{*}$. Fixed values for $r$ are obtained from a fit with a constant function to the $r$ distribution which yields $r = 5.64 \pm 0.05$ ($6.40 \pm 0.06$) for decay mode \modea\ (\modeb). A non-constant $1^{st}$ order polynomial can be fitted to the $r$ distribution for mode \modea; this scenario is investigated in Section~\ref{ssec:fitproc}. The fits to $\Delta m_{res}$ are repeated with $r$ fixed to the constant value obtained which leaves $\sigma_{0}$ as the only free parameter for the width. For each $p^{*}$ bin, $\sigma_0$ is recalculated. The new $\sigma_{0}$ distribution can be best parameterized by the second order polynomial below with coefficients $b_{i}$.
\begin{equation}
\sigma_{0}(p^{*}) = b_{0} +  b_{1}p^{*} + b_{2}p^{*2}.
\label{eq:sigma0}
\end{equation}

\subsection{Validation of the resolution model}
\label{sub:val}
\begin{figure}[!htb]
\begin{center}
\includegraphics[height=5cm]{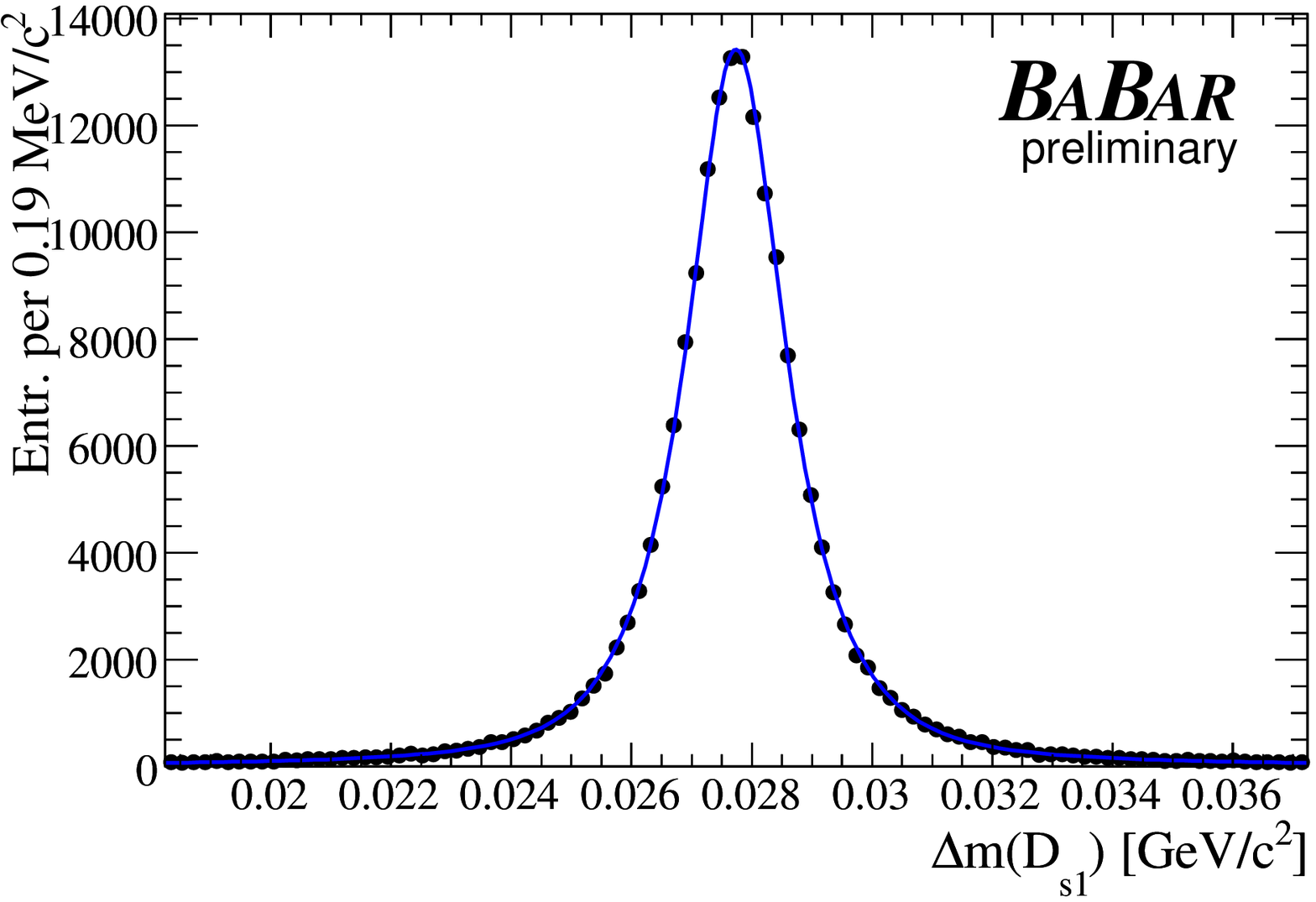}
\includegraphics[height=5cm]{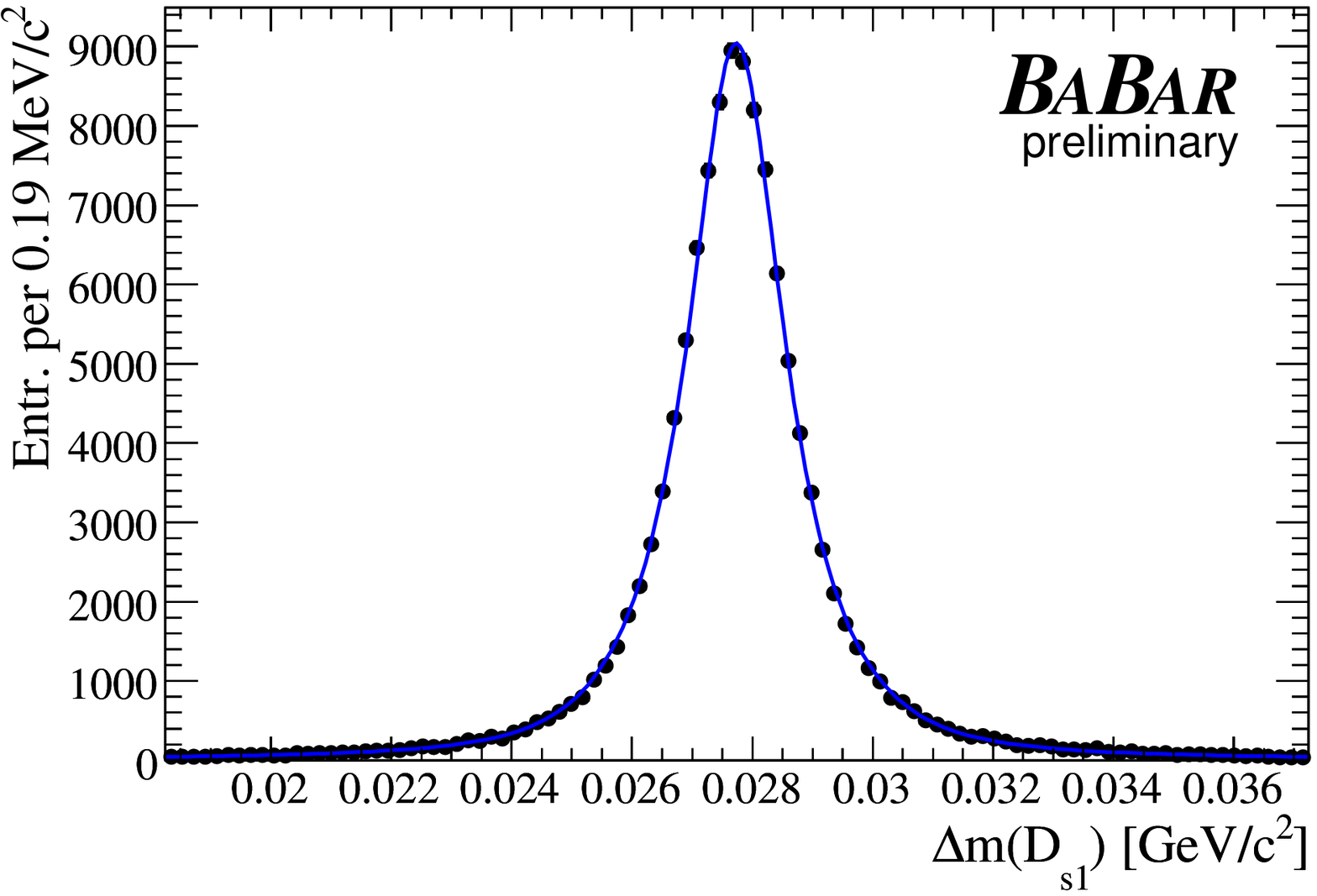}
\caption{Fit of the convolution of the non-relativistic Breit-Wigner function and the resolution function to the \deltadso distribution in MC (dots) as a crosscheck for the $p^*$-dependent resolution model. Left: decay mode \modea; right: decay mode \modeb}
\label{fig:mcfit}
\end{center}
\end{figure}

To verify that the resolution model is reliable, we applied it to the \deltadso distribution of the \Dsop Monte Carlo data sample
\begin{equation}
\deltadso = S(\dmmean,\Gamma(\Dsop)) \ast R(\Delta m_{res}).
\end{equation}
The generated \Dsop mass difference distribution $S(\dmmean,\Gamma(\Dsop))$ in the simulation follows a non-relativistic Breit-Wigner distribution with mean value \dmmean and width $\Gamma(\Dsop)$ (Table~\ref{tab:mcfit}). The fit of this Breit-Wigner convoluted with the resolution function to Monte Carlo data must return the generated values for the mass difference \dmmean and the decay width $\Gamma(\Dsop)$. Since the convolution of the resolution model and the Breit-Wigner function cannot be handled analytically one has to use numerical integration methods instead. To compute the
convolution integral the {\it Trapezoid Sum Rule} method has been
applied. The convolution window has been chosen as $\pm10\mevcc$, which
is about $10$ times the width of the resolution function, and was
divided into 200 bins. The applied method returns stable fit results
for reconstructed Monte Carlo and reproduces the input values for the
mass difference and width in the simulation with sufficient accuracy (Fig.~\ref{fig:mcfit}, Table~\ref{tab:mcfit}). The differences between the reconstructed values and the generated values are assigned as systematic uncertainties, which are $-7\kevcc$ ($-14 \kevcc$) for the mass difference and $-2\kevcc$ ($-9\kevcc$) for the width . 
\begin{table}[!htb]
\caption{Results of the test fits to the MC data (statistical errors only) for both decay modes, compared with the generated values.}
\begin{center}
\begin{tabular}{|c|c|c|c|} \hline
Parameter & \modea & \modeb & generated\\ \hline\hline
$\dmmean\ /\mevcc$ & $27.737 \pm 0.003$ & $27.730 \pm 0.003$ & $27.744$ \\ \hline
$\Gamma(\Dsop)\ /\mevcc$ & $0.998 \pm 0.005$ & $0.991 \pm 0.007$ & $1.000$ \\ \hline
\end{tabular}
\end{center}
\label{tab:mcfit}
\end{table}

\subsection{Fit to the data}
\label{sec:ffit}
The assumption that the \Dsop lineshape $S(\dmmean,\Gamma(\Dsop))$ follows a non-relativistic
Breit-Wigner, as used for the MC, is not sufficient for the measurement of the intrinsic
width. A better and commonly used description of a resonance lineshape
is the following, {\it relativistic} Breit-Wigner function:
\begin{equation}
BW(m)\propto \frac{m m_0 \Gamma}{\left(m^2-m_0^2\right)^2+m_0^2\Gamma^2}
\label{eq_relbw}
\end{equation}

\begin{figure}[!htb]
\begin{center}
\includegraphics[height=6cm]{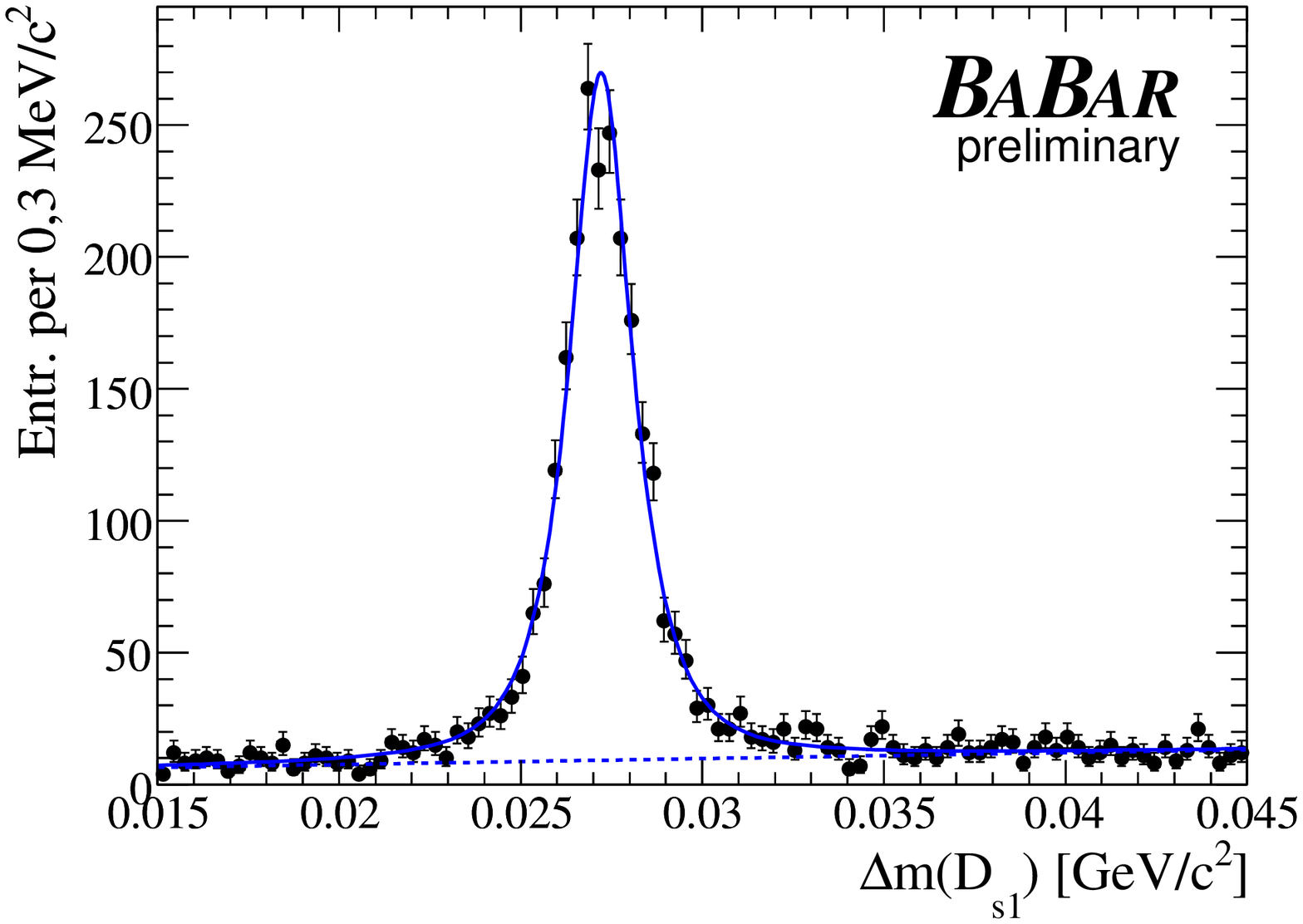}
\includegraphics[height=6cm]{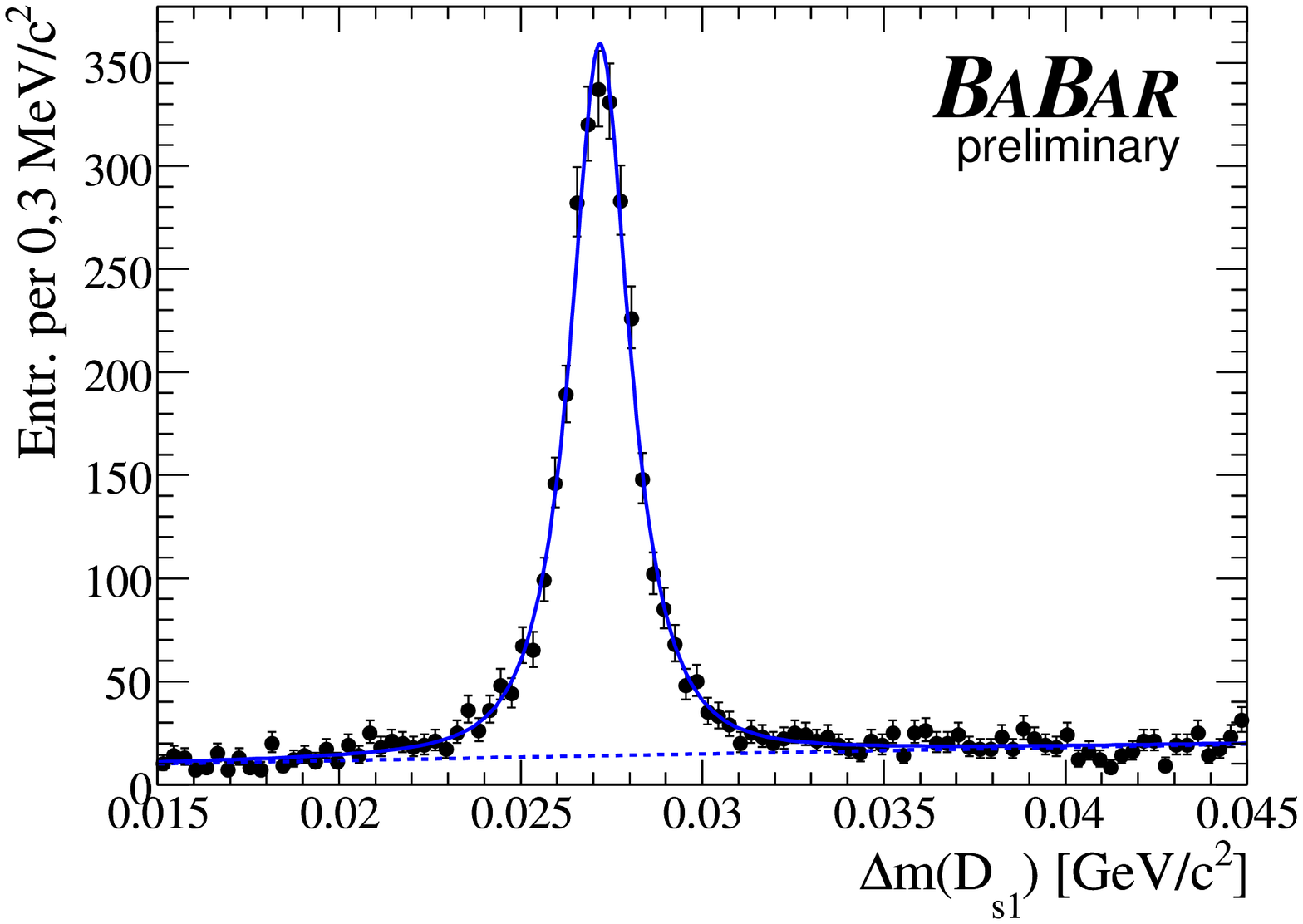}
\caption{Fit to the data (dots) of the relativistic Breit-Wigner function convoluted with the $p^*$-dependent resolution function to the \deltadso spectrum to obtain the mass difference \dmmean and width $\Gamma(\Dsop$). The background is described by a first order polynomial, shown by the dotted line. Left: decay mode \modea; right: decay mode \modeb.}
\label{fig:datfit}
\end{center}
\end{figure}

To measure the mass difference \dmmean and the intrinsic width $\Gamma(\Dsop)$, the extracted $p^{*}$-dependent resolution model $R(\Delta m_{res})$ is convoluted with the relativistic Breit-Wigner distribution and then fitted to data using an unbinned maximum likelihood fit. For the convolution the same numerical integration method as described for the MC test fit (Section~\ref{sub:val}) is applied.
The background is described by a linear function. The fit parameters obtained for both decay modes are listed in table~\ref{tab:datfit}. The corresponding mass distributions are shown with the fits in Fig.~\ref{fig:datfit}. 

\begin{table}[!htb]
\caption{Results of the fit to the data (statistical errors only) for both decay modes.}
\begin{center}
\begin{tabular}{|c|c|c|} \hline
Parameter & \modea & \modeb \\ \hline\hline
$\dmmean\ /\mevcc$ & $27.209 \pm 0.028$ & $27.180 \pm 0.023$ \\ \hline
$\Gamma(\Dsop)\ /\mevcc$ & $1.112 \pm 0.068$ & $0.990 \pm 0.059$ \\ \hline
Signal yield (events) & $2401 \pm 47$ & $2959 \pm 51$ \\ \hline
\end{tabular}
\end{center}
\label{tab:datfit}
\end{table}

\section{SYSTEMATIC UNCERTAINTIES}
\label{sec:Systematics}
	\label{syst}

Sources of systematic uncertainties can be divided into three categories: a) discrepancies between the resolution in data and Monte Carlo (Section~\ref{ssec:systres}); b) extraction of the resolution model and the fit procedure (Section~\ref{ssec:fitproc}); c) uncertainties due to inaccuracies in the track reconstruction (Section~\ref{ssec:detcon}). The results for the different systematic errors are listed in Table~\ref{tab:systematic}. 

\subsection{Resolution studies}
\label{ssec:systres}
A \Dz sample including both decays $\Dz \to K^{-}\pi^{+}$ and $\Dz \to K^{-}\pi^{+}\pi^{+}\pi^{-}$ and a $\KS \to \pi^{+}\pi^{-}$ sample have been extracted from data and generic \ccbar Monte Carlo and are used for resolution studies. The \Dz and \KS mesons have been reconstructed in the same way as described in Section~\ref{ssec:evsel}, but without a requirement for the origin of the particles. 
Due to the negligible width of these two mesons, the resolution can be directly obtained from a fit to the respective signals. To avoid momentum-related discrepancies, the samples are divided into $p^{*}$-dependent subsets. The resolution function from Eq.~\ref{eq:iares} is fitted to each data and \ccbar Monte Carlo subset. The full width at half maximum (FWHM) is calculated for each bin as a measure of the width of the resolution function. For the ratio between the MC and the data resolution, values of $0.90 \pm 0.01$ for the \Dz as well as for the \KS samples are obtained by fitting a constant function to the $p^{*}$-dependent ratio. In order to measure the impact of this deviation on the results for the \Dsop signal lineshape in data, the width of the resolution function obtained in Section~\ref{ssec:resmod} is enlarged by $10\%$. The additional resolution is assumed to follow a Gaussian distribution. To achieve this widening, the $\Delta m_{res}$ spectrum (divided into $p^{*}$ bins as used for the standard fit explained in Section~\ref{ssec:resmod}) is smeared by a Gaussian distribution with a certain width $\sigma_{e}$ which yields a spectrum with a FWHM $10\%$ larger than the original one. The parameters of the resolution function are obtained as described in Section~\ref{ssec:resmod}. With this resolution function, the fit to data is repeated. As expected, the mass is not affected while the width is lowered by $67 \kevcc$ (mode \modea) and $52 \kevcc$ (mode \modeb).

\subsection{Fit procedure}
\label{ssec:fitproc}
Systematic uncertainties concerning the applied method to extract the \Dsop mass and width have been studied.
\begin{paragraph}{Resolution model}
As described in Section~\ref{sub:val}, the resolution model has been tested on the MC data sample. The small differences between the reconstructed values for \dmmean and $\Gamma(\Dsop)$ and the respective generated ones are assigned as systematic errors. 

In a first check, the parameter $r$ of the resolution function is alternatively parameterized by a $1^{st}$ order polynomial. With $r$ now depending on $p^{*}$ rather than being constant, the parameter $\sigma_{0}$ is estimated as described in Section~\ref{ssec:resmod}.
Using the new resolution model, one obtains a negligible shift of the mass and $-25\kevcc$ ($+7\kevcc$) for the decay width.

Another check involves varying the parameter $r$ within one standard deviation $\delta r$ of
its statistical error. Parameterizing $\sigma_0$
using the modified $r$ value, the fit to data is repeated with these resolution parameters. As a conservative estimate the larger value of
the relative change observed for $r-\delta r$ and $r+\delta r$ is taken as an uncertainty. The mass is not affected while the width changes by $-9 \kevcc$ (\modea) and $-27 \kevcc$ (\modeb).
\end{paragraph}
\begin{paragraph}{Numerical integration for convolution}
Lowering and enlarging the integration range by $\pm 1 \mevcc$ while keeping the bin width constant does not yield significant changes. Doubling and halving the bin size while using the standard $20\mevcc$ integration interval yields no relative change of \dmmean and \gammadso either, compared with the standard values.   
\end{paragraph}
\begin{paragraph}{Background parameterization}
In the fit to the \deltadso distribution for data the combinatorial
background was parameterized using a first order polynomial. As a systematic check, the latter is
replaced by a second order polynomial which yields a mass shift of $-3 \kevcc (<0.5 \kevcc)$ and a modification to the width by $-7 \kevcc (-27 \kevcc)$.
In a second test, the background was parameterized by a power-law distribution, e.g. using $BG(\deltadso) = a_{0} + a_{1}\deltadso + a_{2}\deltadso^{a_{3}}$. Both the mass difference and the width do not noticeably change, compared with the standard values. Thus, the deviations from the fit with the $2^{nd}$ order polynomial are taken as the systematic uncertainty arising from the background parameterization.    
\end{paragraph}
\begin{paragraph}{Mass window}
Enlarging the boundaries of the fitted \deltadso
standard region \\($0.015 - 0.045$\mevcc) by $10\mevcc$ has no significant impact on the \dmmean value and changes the width slightly by $+4 \kevcc (-13 \kevcc)$.
\end{paragraph}
\begin{paragraph}{$p^{*}$ correction}
Two strategies - the parameterization and the weighting method - have been followed to take the observed discrepancy between the $p^*$ distributions for data and MC data into account.
 The results obtained from both approaches are consistent within their statistical errors, giving
confidence in the results obtained from the parameterization
method. The deviation between the two methods can be assigned as the
systematic uncertainty arising from the applied correction method, whereby only the width is noticeably affected by $-21 \kevcc\ (-9 \kevcc)$. 
\end{paragraph}

\subsection{Detector conditions}
\label{ssec:detcon}
In the third category of systematic uncertainties, sources arising
from charged particle track reconstruction have been studied. The
amount of material located inside the inner tracking volume (SVT and
DCH) as well as the understanding of the $B$ field inside that volume
plays an important role for the reconstruction of charged
particles. In addition to this, a dependency of the reconstructed mass difference on the $p^{*}$ momentum and on the azimuthal and polar angles $\theta$ and $\phi$ of the $\Dsop$ has been studied.
 
\begin{paragraph}{Material inside tracking volume}
It is possible that the amount of material traversed by charged particles in the tracking volume is underestimated~\cite{ref:bap}. This would result in an incorrect energy loss correction for the tracks.
Two scenarios have been investigated. First, it is assumed that the amount of material inside
the total volume (SVT and DCH) is underestimated by $10\%$. Second, it is assumed that only the amount
of SVT material is underestimated, by $20\%$. To study the effect, the density of the material at the corresponding detector component is increased by the respective number.
With the new detector conditions, the data are reanalyzed and the fit to data is repeated using the standard resolution model with the parameters obtained in Section~\ref{ssec:resmod}. 
As a conservative estimate, the largest deviations of the mass and width obtained from these fits are taken as the systematic uncertainty arising from the tracking region material modification. The results are shown in Table~\ref{tab:systematic}.
\end{paragraph}

\begin{paragraph}{Magnetic Field}
The main component of the magnetic field, which is an essential component for the momentum measurement, is the solenoid field which is itself well measured. More uncertain is the contribution of the solenoid field to the magnetization of the permanent magnets used for final focusing and bending of the beams.     
The magnitude of the solenoid $B$ field is changed by $\pm0.02\%$ for the systematic study. The magnetization is varied by $\pm20\%$ in order to account for
differences between the direct field measurements and the permeability measurements. The largest deviations observed for the rescaled solenoid field and the rescaled magnetization are added in quadrature and used as a conservative estimate of the systematic uncertainty arising from the $B$ field. 
\end{paragraph}

\begin{paragraph}{Length scale}
Another source of uncertainty for the momentum arises from the distance scale. The position of the wires in the DCH is known with a precision of $40 \mum$. With a drift chamber radius of $\approx 40 \cm$, this yields a relative precision of $0.01 \%$. As an estimate for the
uncertainty of the momentum due to the length scale, a systematic error half the size of the uncertainty obtained from the $\pm0.02\%$ variation of the solenoid field is assigned. For the mass difference, this yields a shift of $-5 \kevcc$ ($-8 \kevcc$) for mode \modea\ (\modeb). The width is shifted by $+1 \kevcc$ ($+6 \kevcc$).  
\end{paragraph}

\begin{paragraph}{SVT alignment}
Detector misalignment can affect the measurement of the angles between the tracks and possibly the track momenta. This effect has not yet been studied in this analysis. As a conservative estimate, the results obtained from SVT misalignment studies performed for the $\Lambda_{c}$ mass measurement~\cite{ref:bap} will be used as a systematic uncertainty for the \Dsop mass. This yields an additional error of $\pm 23 \kevcc$.  
\end{paragraph}

\begin{paragraph}{Track quality}
For the standard \Dsop candidate selection as described in Section~\ref{ssec:evsel}, no lower limit has been set for the number of drift chamber hits left by a charged particle. To estimate the effect of improved track quality, the selection of the \Dsop candidates has been repeated, requiring that only those $\pi$ and $K$ candidates are used for the $\Dsop$ reconstruction which have left at least 20 hits in the drift chamber. The determination of the resolution model has been repeated using the new data samples. The mass difference is shifted by $+5 \kevcc$ ($-14 \kevcc$), while the width is shifted by $-105 \kevcc$ ($-92 \kevcc$) with respect to the standard values obtained in Section~\ref{sec:ffit}.
In a second test, the $\pi$ originating from the \Dstarp decay has been excluded from the tighter track selection. Because of its small transverse momentum, it will not traverse large parts of the drift chamber. The mass difference is shifted by $-8 \kevcc$ ($<0.5 \kevcc$), while the width is shifted by $-110 \kevcc$ ($-57 \kevcc$). 
As a conservative estimate, the deviations obtained from the first measurement will be used as a systematic uncertainty. 
\end{paragraph}

\begin{paragraph}{Charge dependence}
The mass difference \dmmean and the width $\Gamma(\Dsop)$ have been determined separately for both charges of the \Dso. The values obtained are consistent within errors with the results for the complete data sample (Table~\ref{tab:depen}). Comparing the different values obtained with the respective standard values yields for the mass difference a $\chi^{2}$ of $1.31$ ($0.34$) for mode \modea\ (\modeb) and for the width a $\chi^{2}$ of $2.08$ ($0.85$).
\end{paragraph}

\begin{paragraph}{Momentum dependence}
The mass difference and the width is measured for five different $p^{*}$ bins in the range between $2.7$ and $4.7 \gevc$. The results obtained do not vary drastically compared with the standard mass difference value obtained from the complete data sample. The weighted mean values for the mass difference and the width lie within the error range of the respective values for the full data sample (Table~\ref{tab:depen}). Comparing the $p^{*}$ distributions with the respective standard values yields for the mass difference a $\chi^{2}$ of $0.76$ ($3.09$) for mode \modea\ (\modeb) and for the width a $\chi^{2}$ of $3.11$ ($0.36$).
\end{paragraph}

\begin{paragraph}{Angular dependence}
A dependency of the reconstructed mass on the azimuthal angle $\phi$ has been observed using the high statistic \Dz data sample created for the resolution studies, while no such effect is visible in the corresponding \ccbar MC. The $\phi$ dependency of the reconstructed \Dz mass follows a sine wave and averages to zero when running over the complete data and thus over the whole $\phi$ range.   
Furthermore, the mass difference \dmmean and the width $\Gamma(\Dsop)$ have been measured in bins of the angle $\theta$ of the \Dsop and in bins of the azimuthal angle $\phi$. In both cases, the values obtained for the different bins are consistent with the values obtained from the fits to the complete data sample. The weighted mean values all lie within the error range of the standard values for \dmmean and $\Gamma(\Dsop)$, respectively (Table~\ref{tab:depen}). Comparing the $\theta$ distributions with the respective standard values yields for the mass difference a $\chi^{2}$ of $5.54$ ($0.97$) for mode \modea\ (\modeb) and for the width a $\chi^{2}$ of $0.98$ ($6.76$). For the $\phi$ distributions one obtains for the mass difference a $\chi^{2}$ of $5.41$ ($2.87$) and for the width a $\chi^{2}$ of $3.28$ ($0.37$). Larger $\chi^{2}$ values are due to bins with greater deviations, caused by the small number of \Dsop candidates available in this range.
\end{paragraph}

\begin{paragraph}{Run dependence}
Finally, the mass difference and the width have been determined separately for different data taking periods. The results obtained for \dmmean and $\Gamma(\Dsop)$ lie within the error range of the standard values for the mass difference and the width, respectively (Table~\ref{tab:depen}). Comparing the different values obtained with the respective standard values yields for the mass difference a $\chi^{2}$ of $0.55$ ($3.36$) for mode \modea\ (\modeb) and for the width a $\chi^{2}$ of $1.16$ ($0.39$).
\end{paragraph}

\begin{paragraph}{}
In summary, for both the mass difference \dmmean and the decay width $\Gamma(\Dsop)$, no significant dependency on the \Dsop momentum, charge and angle or on the time of data-taking have been observed. The fits to the different data subsamples return values within the error range of the results obtained from the fits to the complete data sample.  
\end{paragraph}

\begin{table}
\caption{Summary of systematic uncertainties.}
  \begin{center}
  \begin{tabular}{|l|c|c|c|c|}\hline
    & \multicolumn{2}{c}{\dmmean\ / \kevcc} & \multicolumn{2}{|c|}{$\Gamma(\Dsop)$\ / \kevcc}\\
    & \modea & \modeb & \modea & \modeb \\ \hline\hline
    Resolution width $+10\%$ & $\pm1$ & $<0.5$ & $\pm67$ & $\pm52$ \\\hline
    MC validation & $\pm7$ & $\pm14$ & $\pm2$ & $\pm9$ \\
    Parameterization of $r$  & $<0.5$ & $<0.5$ & $\pm25$ & $\pm7$\\
    Variation of $r$  & $<0.5$ & $<0.5$ & $\pm9$ & $\pm27$\\
    Numerical integration (width)& $<0.5$ & $<0.5$ & $<0.5$ & $<0.5$\\
    Numerical integration (steps)& $<0.5$ & $<0.5$ & $<0.5$ & $<0.5$\\
    Background parameterization  & $\pm3$ & $<0.5$ & $\pm7$ & $\pm27$\\
    Mass window & $\pm1$ & $<0.5$ & $\pm4$ & $\pm13$\\
    $p^*$ correction & $<0.5$ & $\pm1$ & $\pm21$ & $\pm9$\\\hline
    Detector material & $\pm14$ & $\pm24$ & $\pm20$ & $\pm29$\\
    B field          & $\pm10$ & $\pm16$ & $\pm7$ & $\pm13$\\
    Length scale & $\pm5 $ & $\pm8$ & $\pm1$ & $\pm6$ \\
    SVT alignment & $\pm23$ & $\pm23$ & - & - \\
    Track quality & $\pm5$  & $\pm14$   & $\pm105$   &$\pm92$ \\ \hline
    {\bf Quadratic sum} & {\boldmath $\pm31$} & {\boldmath $\pm43$} & {\boldmath $\pm131$} & {\boldmath $\pm119$} \\\hline
      \end{tabular}
   \label{tab:systematic}
  \end{center}
\end{table}

\begin{table}
\caption{Measurements of \dmmean and $\Gamma(\Dsop)$ in dependence of several variables. The first row shows the results from the fits to the full data sample, followed by the charge dependent measurements. Lines 4 to 6 list the weighted mean values obtained from $p^{*}$, $\theta$ and $\phi$ depending measurements. The last two rows contain the results for different data taking periods.}
  \begin{center}
  \begin{tabular}{|l|c|c|c|c|}\hline
    & \multicolumn{2}{c}{\dmmean\ / \mevcc} & \multicolumn{2}{|c|}{$\Gamma(\Dsop)$\ / \mevcc}\\
    Dependency & \modea & \modeb & \modea & \modeb \\ \hline\hline
    {\bf Full data} & {\boldmath $27.209 \pm 0.028$} & {\boldmath $27.180 \pm 0.023$} & {\boldmath $1.112 \pm 0.068$} & {\boldmath $0.990 \pm 0.059$} \\
    $\Dsop$ & $27.176 \pm 0.041$ & $27.193 \pm 0.033$ & $1.201 \pm 0.101$ & $1.042 \pm 0.086$  \\
    $\Dsom$ & $27.239 \pm 0.037$ & $27.166 \pm 0.033$  & $1.007 \pm 0.092$ & $0.933 \pm 0.082$ \\
    $p^{*}(\Dsop)$ & $27.196 \pm 0.073$ & $27.204 \pm 0.058$  & $1.222 \pm 0.188$ & $0.932 \pm 0.155$ \\  
    $\theta(\Dsop)$ & $27.211 \pm 0.053$ & $27.181 \pm 0.045$  & $1.107 \pm 0.129$ & $0.996 \pm 0.116$ \\ 
    $\phi(\Dsop)$ & $27.206 \pm 0.062$ & $27.179 \pm 0.053$  & $1.108 \pm 0.151$ & $0.992 \pm 0.133$ \\ 
    $Run1+2$ & $27.236 \pm 0.041$ & $27.232 \pm 0.036$  & $1.022 \pm 0.098$ & $0.947 \pm 0.092$ \\ 
    $Run3+4$ & $27.188 \pm 0.039$ & $27.145 \pm 0.031$  & $1.167 \pm 0.098$ & $0.958 \pm 0.078$ \\ \hline
    \end{tabular}
   \label{tab:depen}
  \end{center}
\end{table}

\subsection{Techniques for combining results}
The results obtained for the two decay modes are combined using an expanded $\chi^{2}$ method that allows different correlations between the individual components of the systematic uncertainties.  
As a crosscheck, the results are combined using a Best Linear Unbiased Estimate (BLUE) technique \cite{ref:blue} which takes correlated and uncorrelated errors separately into account. The results obtained from the methods are consistent with each other. Since the BLUE method delivers only one single combined error, the results from the first method are reported.

\section{SUMMARY}
\label{sec:Summary}
We have presented a high precision measurement of the mass and the decay width
of the meson $\Dso(2536)^{+}$ using the decay mode $\Dsop \rightarrow \Dstarp\KS$. The mass difference between \Dsop and $\Dstarp\KS$ for the two reconstructed decay modes is measured to be 
\begin{center}
$\dmmean_{K4\pi} = 27.209 \pm 0.028 \pm 0.031 \mevcc$,\\
$\dmmean_{K6\pi} = 27.180 \pm 0.023 \pm 0.043 \mevcc$,\\
\end{center}
with the first error denoting the statistical uncertainty and the second one the systematic uncertainty.
These results correspond to a relative error of $0.15\%$ for the mass difference. This lies within the range of precision achievable with the \babar\ detector: the \jpsi mass has been reconstructed with a relative error of $0.05\%$~\cite{ref:babar}. 

Combining the results, while taking the systematic errors including the uncertainties of the \Dstarp mass ($\pm 0.4\mevcc$) and of the \KS mass ($\pm 0.022 \mevcc$) into account, yields a final value for the \Dsop mass of
\begin{center}
$m(\Dsop) = 2534.85 \pm 0.02 \pm 0.40 \mevcc$,\\ 
\end{center}
while the PDG value for the mass is given as $2535.35 \pm 0.34 \pm 0.50 \mevcc$.
The error on the measured \Dsop mass is dominated by the uncertainty of the \Dstarp mass. The mass
difference between the \Dsop and the \Dstarp follows from these results as
\begin{center}
$\Delta m = m(\Dsop) - m(\Dstarp) = 524.85 \pm 0.02 \pm 0.04 \mevcc$.\\
\end{center}
The decay width is measured to be
\begin{center}
$\Gamma(\Dsop)_{K4\pi} = 1.112 \pm 0.068 \pm 0.131 \mevcc$,\\
$\Gamma(\Dsop)_{K6\pi} = 0.990 \pm 0.059 \pm 0.119 \mevcc$.\\
\end{center}
The final combined value for decay width is
\begin{center}
$\Gamma(\Dsop) = 1.03 \pm 0.05 \pm 0.12 \mevcc$.\\
\end{center}
The result for the mass difference $\Delta m = m(\Dsop) - m(\Dstarp)$ represents an improvement in precision by a factor of 14 compared with the current PDG value of $525.3 \pm 0.6 \pm 0.1 \mevcc$. Our result deviates by $1\sigma$ from the larger PDG value. The precision achieved is comparable with other recent high precision analyses performed at \babar\, like the $\Lambda_{c}$ mass measurement (\mbox{$m(\Lambda_{c}) = 2286.46 \pm 0.04 \pm 0.14 \mevcc$)~\cite{ref:bap}}.
Furthermore, this analysis presents for the first time a direct measurement of the \Dsop decay width with small errors rather than just an upper limit, which is currently stated by the PDG as $2.3 \mevcc$.

\section{ACKNOWLEDGMENTS}
\label{sec:Acknowledgments}

We are grateful for the 
extraordinary contributions of our \pep2\ colleagues in
achieving the excellent luminosity and machine conditions
that have made this work possible.
The success of this project also relies critically on the 
expertise and dedication of the computing organizations that 
support \babar.
The collaborating institutions wish to thank 
SLAC for its support and the kind hospitality extended to them. 
This work is supported by the
US Department of Energy
and National Science Foundation, the
Natural Sciences and Engineering Research Council (Canada),
Institute of High Energy Physics (China), the
Commissariat \`a l'Energie Atomique and
Institut National de Physique Nucl\'eaire et de Physique des Particules
(France), the
Bundesministerium f\"ur Bildung und Forschung and
Deutsche Forschungsgemeinschaft
(Germany), the
Istituto Nazionale di Fisica Nucleare (Italy),
the Foundation for Fundamental Research on Matter (The Netherlands),
the Research Council of Norway, the
Ministry of Science and Technology of the Russian Federation, and the
Particle Physics and Astronomy Research Council (United Kingdom). 
Individuals have received support from 
the Marie-Curie IEF program (European Union) and
the A. P. Sloan Foundation.

\end{document}